\documentclass[aps,pra,superscriptaddress,showpacs,reprint]{revtex4-1}

\usepackage{hyperref, soul}
\usepackage{graphicx}
\usepackage{amsfonts,amsmath,amssymb}
\usepackage{appendix}
\usepackage[latin1]{inputenc}
\usepackage{color}
\usepackage{ulem}
\definecolor{forestgreen}{RGB}{34, 139, 34}

\begin{document}

\title{Rotation of the noise ellipse for squeezed vacuum light generated via four-wave-mixing}

\author{Neil V. Corzo}
\affiliation{Joint Quantum Institute, National Institute of Standards and Technology and the University of Maryland, Gaithersburg, Maryland 20899, USA.}
\affiliation{Center for Photonic Communication and Computing, EECS Department,
Northwestern University, 2145 Sheridan Road, Evanston, Illinois 60208-3118, USA.}
\author{Quentin Glorieux}
\email{quentin.glorieux@nist.gov}
\affiliation{Joint Quantum Institute, National Institute of Standards and Technology and the University of Maryland, Gaithersburg, Maryland 20899, USA.}
\affiliation{Group of Applied Physics, University of Geneva, Chemin de Pinchat 22, CH-1211 Geneva, Switzerland}
\author{Alberto M. Marino}
\affiliation{Homer L. Dodge Department of Physics and Astronomy, The University of Oklahoma, 440 W. Brooks St., Norman, Oklahoma 73019, USA.}
\author{Jeremy B. Clark}
\affiliation{Joint Quantum Institute, National Institute of Standards and Technology and the University of Maryland, Gaithersburg, Maryland 20899, USA.}
\author{Paul D. Lett}
\email{paul.lett@nist.gov}
\affiliation{Joint Quantum Institute, National Institute of Standards and Technology and the University of Maryland, Gaithersburg, Maryland 20899, USA.}

\pacs{42.50.Dv, 42.50.Lc, 42.50.Nn}

\begin{abstract} 
We report the generation of a squeezed vacuum state of light whose noise ellipse rotates as a function of the detection frequency.
The squeezed state is generated via a four-wave mixing process in a vapor of $^{85}$Rb.
We observe that rotation varies with experimental parameters such as pump power and laser detunings.
We use a theoretical model based on the Heisenberg-Langevin formalism to describe this effect.
Our model can be used to investigate the parameter space and to tailor the ellipse rotation in order to obtain an optimum squeezing angle, for example, for coupling to an interferometer whose optimal noise quadrature varies with frequency.

\end{abstract}

\maketitle

\section{Introduction}
The generation of non-classical states of light has been a topic of interest since the first experimental demonstration of squeezed  light by Slusher et al. using four-wave mixing (4WM) in sodium vapor \cite{Slusher:1985vm}.
More recently interest in 4WM techniques has seen a resurgence due to the potential for the generation of very high multi-spatial modes \cite{Boyer:2008ts} quantum correlations (two-mode twin beams) in the so-called phase-insensitive configuration \cite{McCormick:07,Glorieux:2010iv,Liu:11,Qin:12,Glorieux:2012km}.
4WM in the phase-sensitive configuration can also produce single-mode vacuum squeezed states \cite{Corzo:11} that might be used for improving sensitivity in precision interferometry.

Gravitational wave detection using optical interferometers with long arms is a promising but highly challenging technique \cite{Schnabel:2010cr}.
Since a pioneering experiment with  modest sensitivity \cite{Forward:1978iv}, the performance of interferometers has improved dramatically \cite{Abadie:2011dj,Vahlbruch:2010bp,Lastzka:2008wd}.
In these experiments, gravitational waves would be detected as a phase or path difference between the two arms of the interferometer and the sensitivity is  ultimately limited by the quantum noise.
The standard quantum limit (SQL) has been introduced for this class of interferometers \cite{Braginsky:1996tr}, and it is equivalent to the detection limit for the arm length difference: 
\begin{equation}
\Delta_{SNL}=\sqrt{\frac{\hbar \tau}{m}},
\end{equation}
where $\tau$ is the measurement time and $m$ is the mass of the mirrors.
In 1980, Caves emphasized that the SQL in interferometers is a property of the interferometer itself and it is a consequence of the vacuum noise entering from the unused port \cite{Caves:1980dw}.
Subsequently, he proposed to inject squeezed light into the unused port to improve the sensitivity of these interferometers \cite{Caves:1981vh}, which has been experimentally demonstrated \cite{Grangier:1987wo,Xiao:1987vo} and is currently used on a daily basis in the LIGO project \cite{Abadie:2011dj}.
This technique, however, cannot overcome the SQL at all frequencies because the reduction in photon noise at low frequencies is overwhelmed by an increase in the random motion of the mirrors known as radiation pressure fluctuations.
In an extension of Caves idea, Unruh demonstrated that the SQL can be violated across a large band of frequencies if the appropriate quadrature component of the mode entering the unused port is squeezed \cite{Meystre:1983ww}.
The key feature of the Unruh proposal is that the squeezed vacuum angle (equivalent to the direction of the noise ellipse) be frequency dependent.
Since the dominant source of noise at low frequencies is the radiation pressure noise, and at high frequencies is the photon shot noise, this determines that the choice of the noise ellipse direction where the squeezing is present should rotate between the phase quadrature at low frequencies and the amplitude quadrature at high frequencies.
Several papers have discussed this proposal \cite{Jaekel:2007uu,Pace:1993uq} and experimental demonstrations have been achieved \cite{Kimble:2001bz}.
Different techniques have been proposed to implement a frequency dependent noise ellipse \cite{Khalili:2009va,Corbitt:2004us,PhysRevA.73.053810}
Most of these techniques are based on pre-filtering cavities as initially proposed in \cite{Kimble:2001bz}, and they can all introduce losses and mode-matching problems.
More recently, Horrom et al. have demonstrated an example of angle rotation of the quantum noise quadratures using the frequency-dependent absorption of the EIT window\cite{horrom13}.\\

In this paper we discuss a different technique that allows for the rotation of the noise ellipse of vacuum squeezed light using  four-wave mixing (4WM) in a hot atomic vapor.
In this technique, we use a double-$\Lambda$ system \cite{Glorieux:2011fj} in a phase-sensitive configuration to generate squeezed vacuum light, described in detail in \cite{Corzo:11}.
We demonstrate up to  $-4.0$~dB of squeezing bellow the shot noise limit (SNL) and we show that the squeezing spectrum can be modified to maximize squeezing at various analysis frequencies. 
We propose an explanation for this effect, based on the fact that two frequency sidebands (positive and negative sidebands) contribute differently to the measured noise, and we verify the consistency of this description with our experimental data.
We report an experimental observation of the noise ellipse rotation over a frequency bandwidth of several MHz.
We suggest that the rotation of the noise ellipse in this system is due to a frequency dependent phase shift induced by the non-linear susceptibility of the atomic medium.
We adapt the theoretical model of \cite{Glorieux:2010ja} to our system in order to provide a quantitative estimation of the the phase shift.
For the experimental parameters used in this paper we observe a noise ellipse rotation of $\pi/5$ over 10~MHz frequency bandwidth.
This combination of experimental results and theoretical model might allow one to tailor the noise ellipse rotation to follow the optimal squeezing angle specific to a given interferometer. 
\\

\section{Experimental setup}
\begin{figure}[]
\centering\includegraphics[width=1\columnwidth]{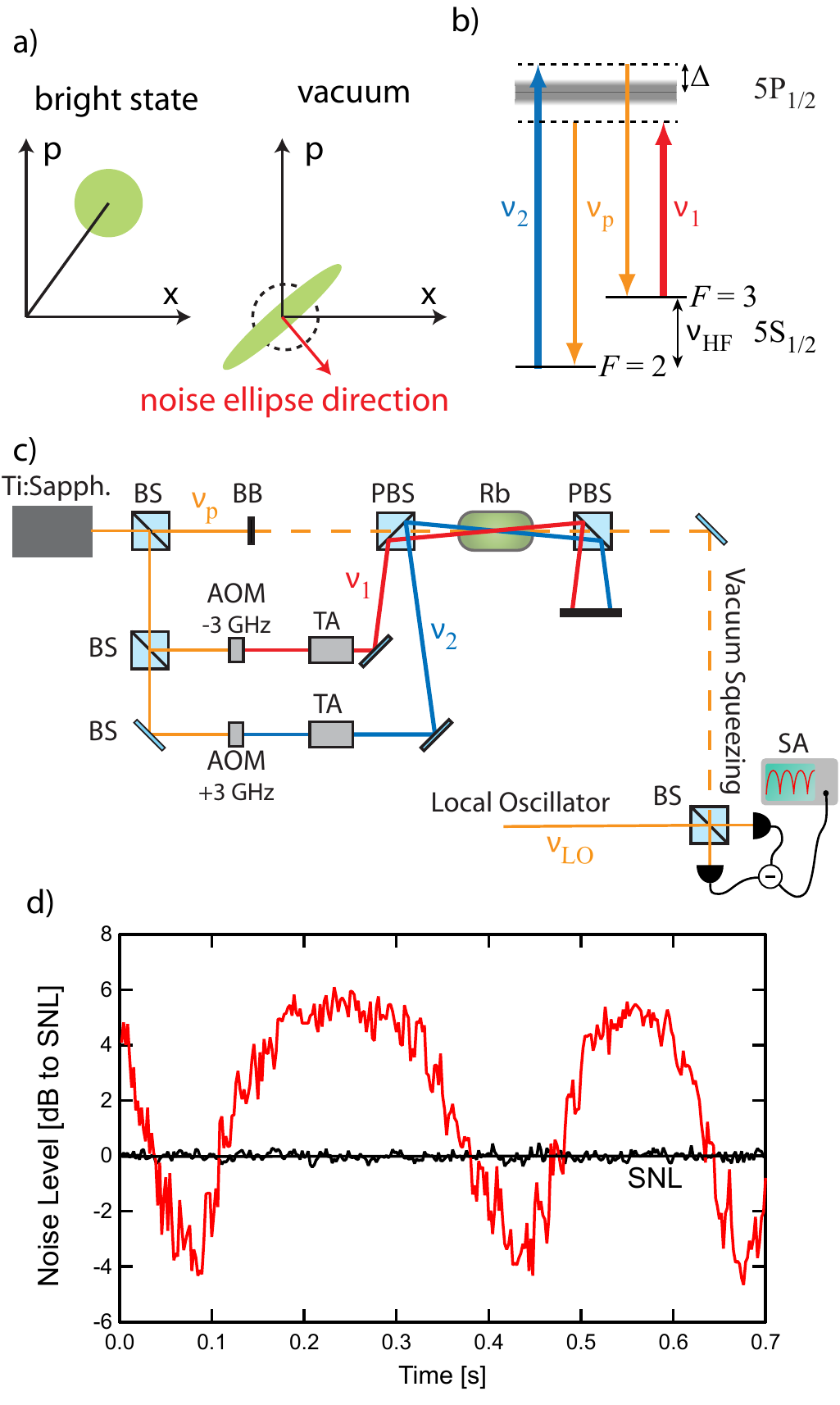}
\caption{\label{fig1} (Color online.) a) Ball-and-stick representation for bright beams and vacuum states.
b) Energy levels of the D1 line in $^{85}$Rb and the optical frequencies arranged in the double-$\Lambda$ system. c) Experimental setup to generate squeezed light and measure the squeezing level. AOM: acousto-optic modulator, TA: tapered amplifier, SA: radio-frequency spectrum analyzer, BS: non-polarizing beamsplitter, PBS: Polarizing beamsplitter, BB: beam block. d) Typical vacuum squeezing noise measurement at 1~MHz analysis frequency (red/gray) compared to the shot noise limit (black).}
\end{figure}

A common representation of the fluctuations of a quantum state as a phase space diagram is known as the "ball-and-stick" picture \cite{Bachor:2004uk}. 
In this representation, the stick corresponds to the mean value of the field and the ball corresponds to the area of the fluctuations around the mean value.
For a vacuum state, the mean value is zero and therefore the stick disappears and the ball is centered around the origin (see  Fig.~1~a).
The standard quantum limit in this representation is described by a circle with a diameter normalized to 1.  
A squeezed vacuum state is therefore an ellipse with one direction for which the axis is smaller than 1, referred to as the noise ellipse direction (see Fig.~1(a)). In an ideal case, the area of the fluctuation ellipse is preserved, otherwise the area is increased.
In the following, we study the compression factor of this ellipse (referred to as the squeezing value) and the angle of this ellipse generated by the four-wave mixing process in $^{85}$Rb.

The experimental setup is similar to the one presented in \cite{Corzo:11}  and is shown in Fig.~\ref{fig1}(c). 
A Ti:sapphire laser at frequency $\nu_{p}$ is tuned approximately $0.8$~GHz to the blue of the $^{85}$Rb 5S$_{1/2}, F=3\rightarrow$ 5P$_{1/2}$ transition (red of the $^{85}$Rb 5S$_{1/2}, F=2\rightarrow$ 5P$_{1/2}$ transition), as shown in Fig.~\ref{fig1}(b).
The one-photon detuning is referred to as $\Delta$.
 Two pump frequencies are generated by frequency-shifting the light of the Ti:sapphire laser using two double-passed 1.52~GHz acousto-optic modulators (AOMs). 
One pump beam (frequency $\nu_{1}$) is downshifted  and the other upshifted (frequency $\nu_{2}$) by the same amount, around $3$ GHz, that is related to the ground state hyperfine splitting $\nu_{HF}=3036$~MHZ in $^{85}$Rb. 
The beams out of the AOMs (0.5~mW each) are used to seed two tapered amplifiers and generate two strong ($\sim$500~mW each) pump beams. 
The pump beams are spatially-filtered with polarization-maintaining fibers and we obtained maximum powers of 230~mW and 190~mW for the pump beams at frequencies $\nu_{1}$ and $\nu_{2}$, respectively.
The two pumps have the same linear polarization which are orthogonal to the polarization of the probe. 
These pump beams are directed into a 12.5~mm long vapor cell filled with isotopically pure $^{85}$Rb heated to 90~$^{\circ}$C and with an angle of 0.4~degrees between the beams, as shown in Fig.~\ref{fig1}(b).
The pumps are focused at the center of the cell with waists ($1/e^{2}$ power radius)  of 600~$\mu$m.  
After the cell the pump light is filtered out using a Glan-Taylor polarizer and the output squeezed vacuum beam is sent to a balanced homodyne detection system and the noise is then measured using a RF spectrum analyzer. \\

The local oscillator (LO) for homodyne detection, at frequency $\nu_{LO}$, is taken directly from the Ti:sapphire laser. 
The LO carrier frequency selects the frequency of the generated photons that are being analyzed. 
We summarize in Fig.~2~(a), the various frequencies involved.   
The two photon detuning $\delta$ is defined as $\delta = \nu_{LO} - (\nu_{1} + \nu_{HF})$ and is also equivalent to $\nu_{LO}-\nu_p$ for the measurements presented here.
Fig.~1(d) shows a typical measurement of $-4.0$ dB of vacuum quadrature squeezing at $\Delta=0.8$~GHz and $\delta=4$~MHz for an analysis frequency $\omega_a=$~1~MHz.

As in \cite{Corzo:11}, the homodyne detection system can be locked to measure the quadrature with the minimum noise level of the squeezed vacuum  beams. 
The locking system \cite{McKenzie:05} works as follows: the phase of the squeezed beam is modulated at a frequency of 5 kHz, which translates into a modulation of the noise power signal.  
This signal is then detected at an analysis frequency of 1~MHz (zero span) and  demodulated with the use of a lock-in amplifier. 
As a result we obtain an error signal that has a zero crossing at the minimum and maximum noise levels, which allows us to lock the homodyne detection system and  measure the squeezing and anti--squeezing levels of the output squeezed beam without phase fluctuations.
The main result of this paper is the comparison between measurements with and without a locking system for the phase of the LO.

\begin{figure}[]
\centering\includegraphics[width=0.75\columnwidth]{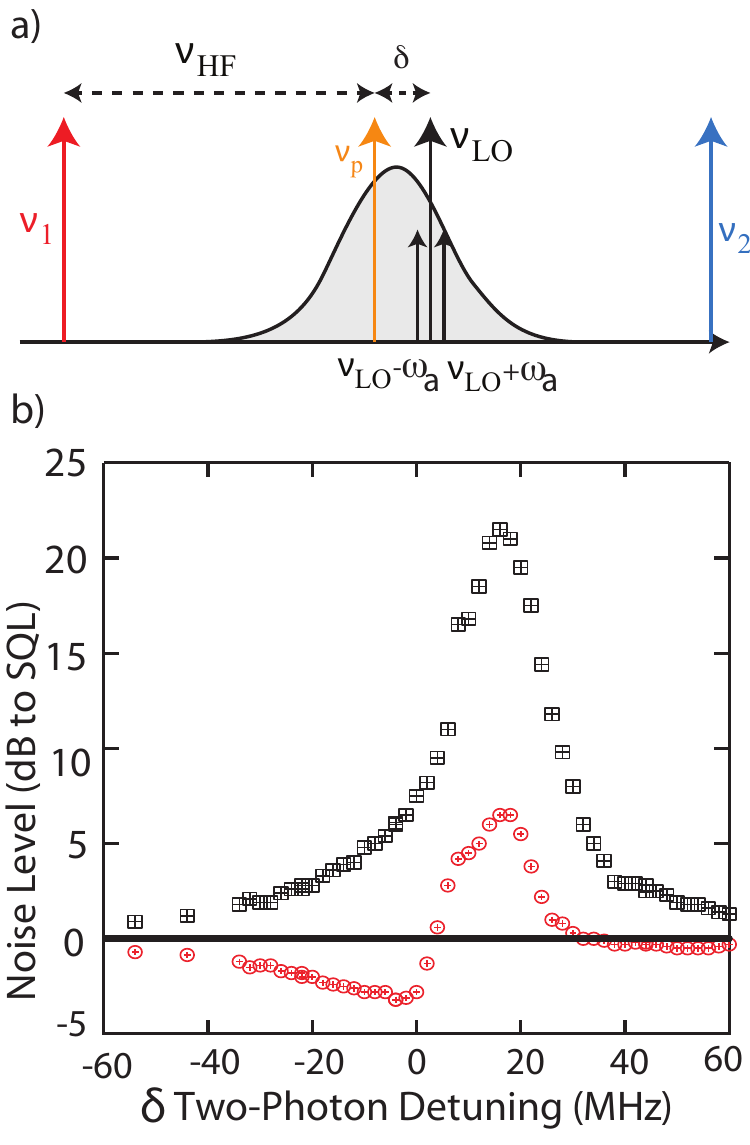}
\caption{\label{fig2}  (Color online.)  a) Diagram to summarize the various frequencies involved. $\nu_1$ and $\nu_2$  are the carrier frequencies of pumps 1 and 2, respectively.  $\nu_{LO}$ is the carrier frequency of the local oscillator and $\nu_p=\nu_1+\nu_{HF}$. The two-photon detuning $\delta=\nu_{LO}-\nu_{p}$. The analysis frequency is denoted $\omega_a$, and the upper and lower sidebands are at $\nu_{LO}\pm\omega_a$.  The grey area represents the gain region of the process (4WM bandwidth). b) Measured noise level of the generated beam as a function of the two-photon detuning $\delta$ at an analysis frequency of $\omega_a=1$~MHz. Black squares corresponds to the choice of phase that maximizes the noise and red circles corresponds to the phase that minimizes the noise. The state is squeezed for a noise level less than zero. Black line: shot noise level. }
\end{figure}

\begin{figure*}[]
\centering\includegraphics[width=1.45\columnwidth]{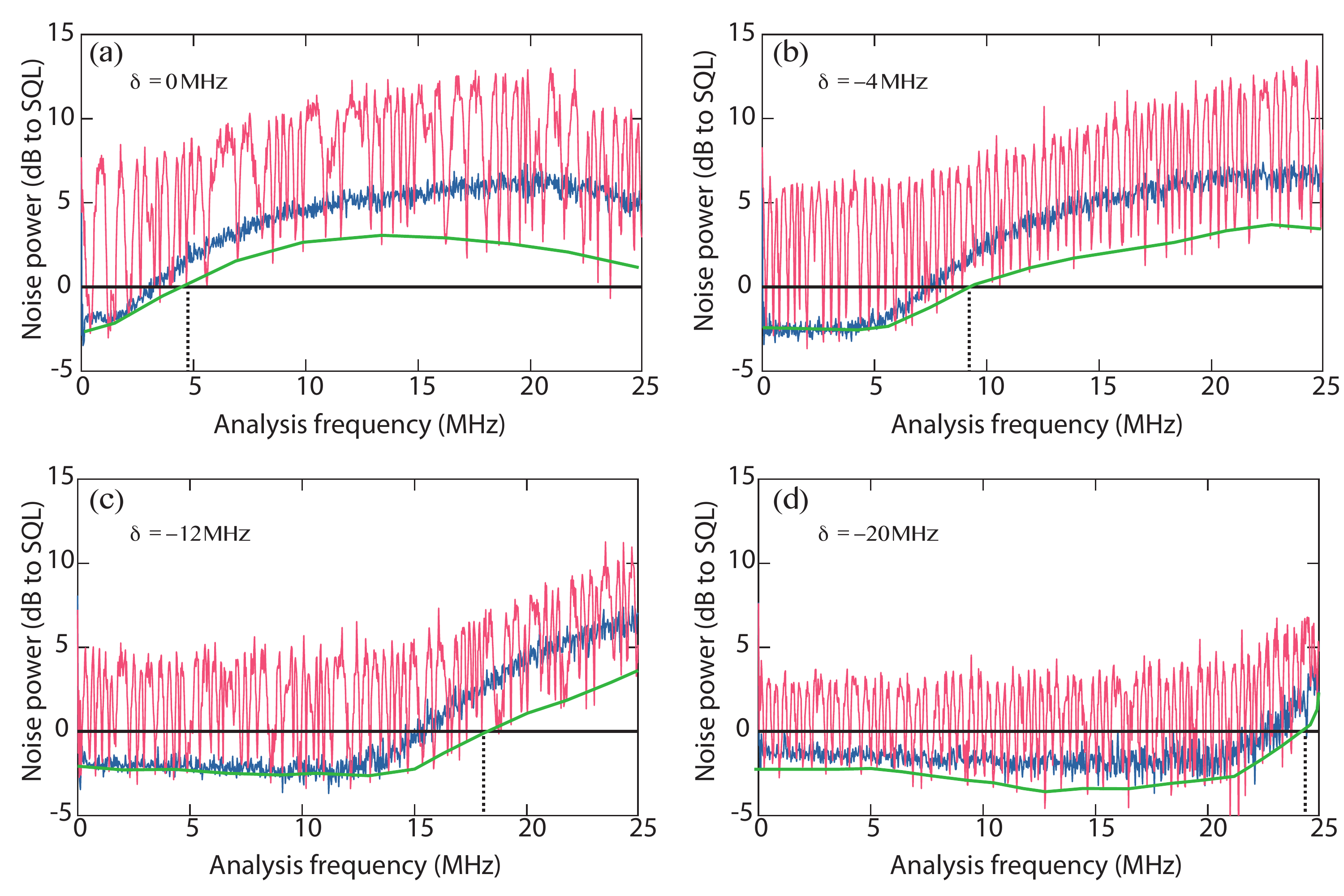}
\caption{\label{fig3}  (Color online.) Measured squeezing as a function of frequency for different values of the two-photon detuning $\delta$. (a)~$\delta=0$~MHz (b)~$\delta=-4$~MHz (c)~$\delta=-12$~MHz (d)~$\delta=-20$~MHz. Red traces (light gray): Measured noise power when the phase of the LO is scanned.
Blue traces (dark gray): Measured noise power when the phase of the LO is locked (at $1$~MHz). 
Green (gray) solid line: envelopes of the detected noise minima, calculated by applying a low pass filter to the red traces.
Black line at 0~dB: shot noise level. $\Delta=0.8$~GHz, the cell temperature $\rm T_{cell}=90^{\circ}C$, and the power in pump 1 is 230~mW and  in pump 2 is 160~mW.}
\end{figure*}

\section{Squeezing bandwidth}

In order to study the noise level on the squeezed beam as a function of the two-photon detuning $\delta$, we show in Fig.~\ref{fig2}(b) the measured noise (minimum and maximum noise) as a function of $\delta$.
We observe the largest squeezing around $\delta=-2$~MHz and the largest excess noise around  $\delta=+18$~MHz.

Similarly, we can measure the noise power for a fixed quadrature phase and two-photon detuning $\delta$ and report it as function of the analysis frequency.
This measurement is known as the squeezing spectrum.
In this section, we clarify the relationship between these two measurements and verify the consistency of the two sets of data.
In Fig. \ref{fig3}, we present the measured squeezing spectrum for different values of the two-photon detuning.
As described earlier, two measurements of the noise power are reported for each value of $\delta$, with (blue/dark gray trace) and without (red/light gray trace) locking the phase of the LO.
In order to quantify the behavior of the noise power as a function of the analysis frequency, we introduce the shot noise limit frequency (SNLF) defined as the analysis frequency where the minimum of the detected noise (red/light gray traces) is equal to the shot noise limit. 
To evaluate this quantity, we calculate the envelope of the detected noise minima, by applying a low pass filter and we plot the resulting filtered curves in Fig.~\ref{fig3}.
These envelopes represent the maximum squeezing obtainable independent of the choice of phase for the LO.
We then report in Table 1 the frequencies at which these envelope curves cross the standard quantum limit.

\begin{table}[h]
\begin{tabular}{ccc}
\toprule
Two-photon detuning &SNLF \hspace{1cm}\\
\colrule
0 MHz &4.9 MHz\\
$-$4 MHz &9.1 MHz \\
$-$12 MHz &17.7 MHz \\
$-$20 MHz & 24.8 MHz \\
\botrule
\end{tabular}

\caption{Behavior of the measured noise power as a function of the two-photon detuning $\delta$. SNLF (shot noise limit frequency) is the analysis frequency at which the squeezing is lost (envelopes of red traces in Fig.\ref{fig3}).}
\label{tab1}
\end{table}

We  note that the values reported on Table 1 show that $4.8$~MHz$<$ SNLF$+\delta$ $<5.7$~MHz, and suggest that SNLF$+\delta$ is a constant within experimental uncertainties.
This point will be addressed in detail at the end of this section.

\begin{figure}[]
\centering\includegraphics[width=0.9\columnwidth]{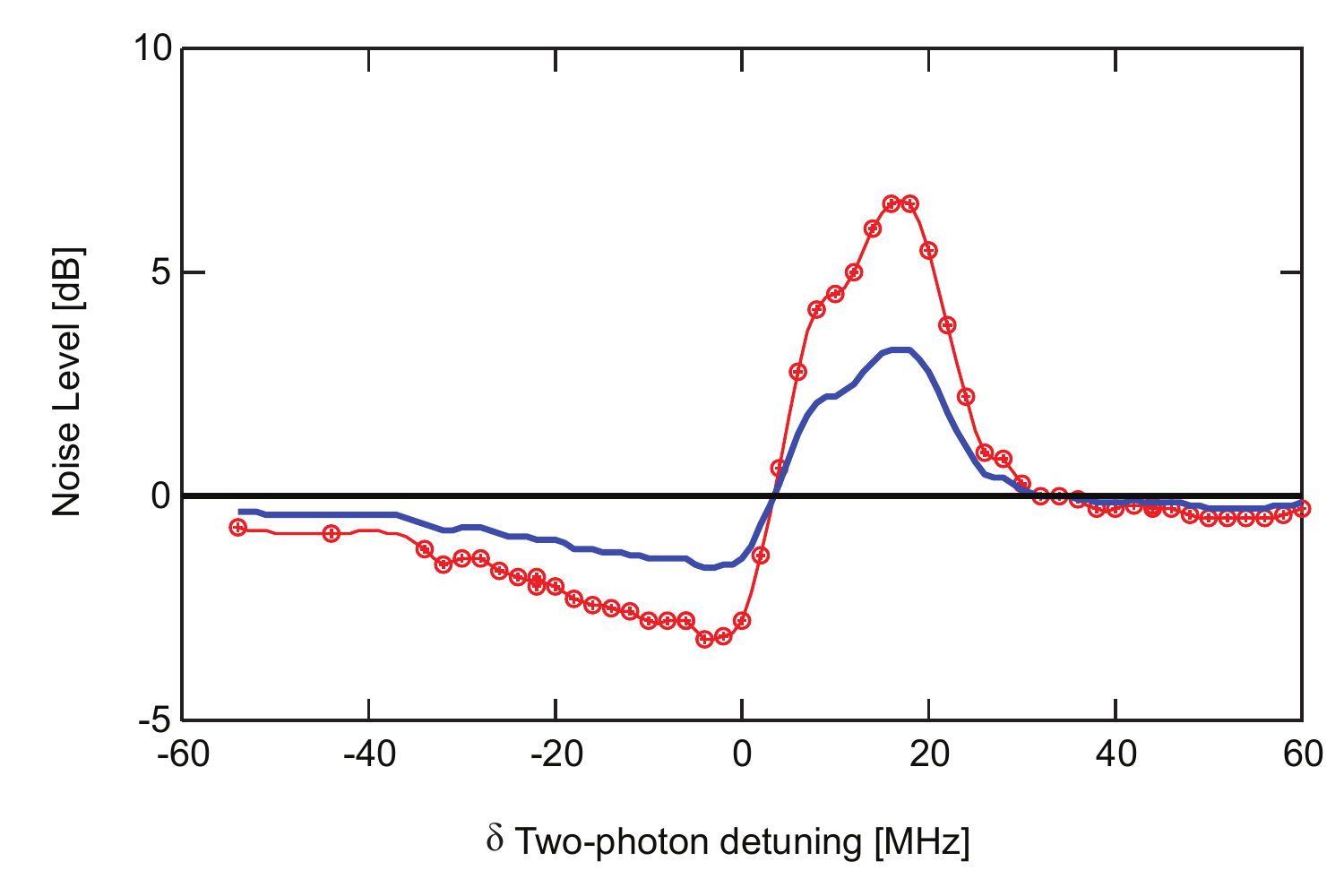}
\caption{\label{fig4} (Color online.) Noise level at an analysis frequency of 0~MHz as a function of the two-photon detuning $\delta$. 
The red circles correspond to the measured noise power as a function of $\delta$ at 1~MHz. 
Blue solid line:  calculated noise at an analysis frequency of 0~MHz as a function of $\delta$. 
Black solid line at 0~dB: shot noise level.}
\end{figure}

When we measure the noise properties of the generated probe beam we beat it with a LO and the frequency of the LO defines the two-photon detuning $\delta$.
The 4WM process, in the phase sensitive configuration used here, spontaneously  generates photons in the spatial mode of the probe beam over a range of frequencies, and the frequency spread is defined as the bandwidth of the process.
The spatial mode is defined by the mode-matching conditions.
In consequence, the LO beats with multiple independent frequencies spontaneously generated in the probe mode, and each of these contributes incoherently to the measured noise spectrum.
The contribution of the upper and lower sidebands must be taken into account separately to describe the noise power at a given analysis frequency $\omega_a$.
More precisely, if we fix the two-photon detuning of the LO ($\delta$), we can write the noise power ($N$) at $\omega_a$ as the sum of the contribution from the two sidebands (see Fig.~2~(a)) for a noise power measured at $\omega_a=0$ with an apparent two photon detuning $\delta+\omega_a$ and  $\delta-\omega_a$.
The noise power can be expressed as follows\cite{perso}:
\begin{equation}
\label{eq1}
N(\omega_a,\delta)=\frac{N(0,\delta+\omega_a)+N(0,\delta-\omega_a)}{2}.
\end{equation}
The noise power at  $\omega_a=0$~Hz is a quantity virtually impossible to measure experimentally.
Nonetheless we can extrapolate its value using the data reported in Fig.~\ref{fig2}, measured at $\omega_a=1$~MHz.
Indeed, the quantity plotted as the red dots in Fig.~\ref{fig2} is $N(\omega_a=1~\text{MHz},\delta)$.
Using Eq.~(\ref{eq1}), we know that :
\begin{equation}
\label{eq2}
N(\omega_a=1~\text{MHz},\delta)=\frac{N(0,\delta+ 1~\text{MHz})+N(0,\delta-1~ \text{MHz})}{2}.
\end{equation}
Estimating the noise power at  $\omega_a=0$~Hz  is then equivalent to solving the following equation:
 \begin{equation}
 	f(\delta)=g(\delta+1)+g(\delta-1),
 \end{equation}
where we want to obtain the function $g(\delta)$ in terms of the function $f(\delta)$.
$f(\delta)$ corresponds to $N(\omega_a=1~\text{MHz},\delta)$, and $g(\delta)$ corresponds to $N(\omega_a=0,\delta)$. 
To solve this equation we make use of the Fourier transform and find the solution:
\begin{eqnarray}
	g(\delta)=\frac{1}{2\pi}\int^{\infty}_{-\infty}\frac{\tilde{f}(t)}{2\cos(2\pi i \delta t)}e^{2\pi i \delta t}dt,
\end{eqnarray}
where $\tilde{f}(t)$ is the Fourier transform of the function $f(\delta)$.
We can then use the data in Fig.~\ref{fig2} to calculate the noise level at an analysis frequency of 0~Hz as function of $\delta$ (see Fig.~\ref{fig4}).

\begin{figure}[]
\centering\includegraphics[width=\columnwidth]{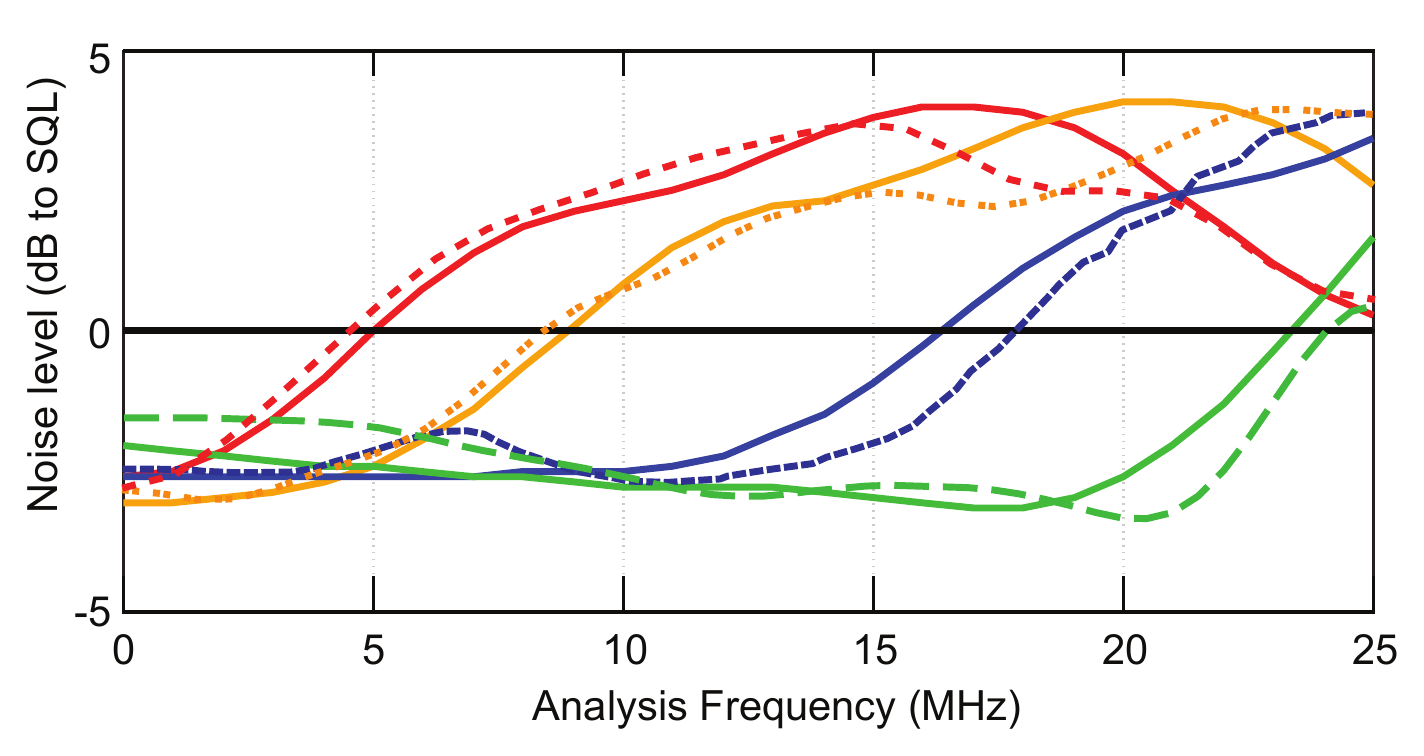}
\caption{\label{fig5}  (Color online.) Noise power as a function of the analysis frequency for $\delta=0$~MHz (red), $-4$~MHz (orange), $-12$~MHz (blue), $-20$~MHz (green).
Dashed lines (regular, double, condensed and large) are the envelopes corresponding to the minimum of noise traces shown in Fig. 3  for respectively $\delta=0$~MHz, $-4$~MHz, $-12$~MHz, $-20$~MHz.
Solid lines are obtained numerically by summing the contributions of the two sidebands using the noise level at  $\omega_a=0$ plotted in Fig.~4.}
\end{figure}

With the estimated noise power  at  $\omega_a=0$~Hz, it becomes straightforward to calculate the expected squeezing spectrum using Eq.~(\ref{eq1}).
In Fig. \ref{fig5}, we plot these calculated traces together with the envelope of the maximum obtainable squeezing (envelope of the oscillating red/light gray traces in Fig. \ref{fig3}).
The qualitative agreement between these two data sets demonstrates the validity of our  approach using the incoherent sum of the two sidebands.
The values of SNLF are also well predicted using the simulated noise traces.
Moreover, we can explain why SNLF$+\delta$ is a constant within the experimental uncertainties.
If one of the sidebands exhibits a much larger excess noise in comparison to the other one, the noise power can be approximated as coming only from the noisiest sideband.
As we can see in Fig.~\ref{fig4}, a sideband at a $\delta$ higher than 1~MHz will exhibit excess noise compared to the SNL, while on the other side ($\delta$ lower than -1~MHz) the noise power in the sideband is below the SNL and its contribution to the total noise power can therefore be neglected.
Therefore the squeezing bandwidth measurements (noise level as a function of $\omega_a$, as in Fig. \ref{fig5}) for values of $\delta\leq 0$ can simply be seen as a translation of the frequency axis by the value of $-\delta$.

\section{Heisenberg Langevin theoretical model}
In this section, we address the question of the discrepancy in Fig. \ref{fig3} between the envelope of the red (light grey) traces representing the minimum obtainable noise independent of the choice of the LO phase indicated along the bottom axis  (referred to as $N_{min}$) and the blue (dark gray) traces representing the noise for a choice of LO phase that minimizes the noise at $\omega=1$~MHz (referred to as $N_{1}$).
As seen in the previous section, a model that sums the contribution of the two frequency sidebands in order to calculate the spectrum of the squeezing can qualitatively explain the global behavior of $N_{min}$.

In order to understand the behavior of  $N_{1}$, we introduce a new parameter: the phase of the LO.
As explained in the experimental section, when we lock the phase of the LO, the reference signal for the lock is derived from the measured and modulated noise at $\omega=1$~MHz.
We denote the phase that minimizes the noise for $\omega=1$~MHz as  $\phi_1$.
We also introduce the phase dependency to the noise power and $N$  becomes a function of three parameters : $\omega_a$, $\delta$, and $\phi$.  
We can use Eq.~(2) to express $N_{min}$: 
 \begin{eqnarray}
 	N_{min}(\omega_a,\delta)&=&\frac12 \min_\phi \left[ N(0,\delta-\omega_a,\phi)\right.\nonumber\\
&&+ \left. N(0,\delta+\omega_a,\phi+\Delta\phi)\right],
 \end{eqnarray}
and $N_{1}$:
 \begin{eqnarray}
	N_{1}(\omega_a,\delta)&=&\frac{\left[ N(0,\delta-\omega_a,\phi_1)+N(0,\delta+\omega_a,\phi_1+\Delta\phi)\right]}{2}.\quad
 \end{eqnarray}
In these equations, $\Delta\phi$ corresponds to an arbitrary phase shift between the two sideband contributions.
This parameter is the main difference with the model discussed in the previous section (where  $\Delta\phi$ was assumed to be 0).
The physical origin of this phase shift will be discussed later.

\begin{figure}[]
\centering\includegraphics[width=0.95\columnwidth]{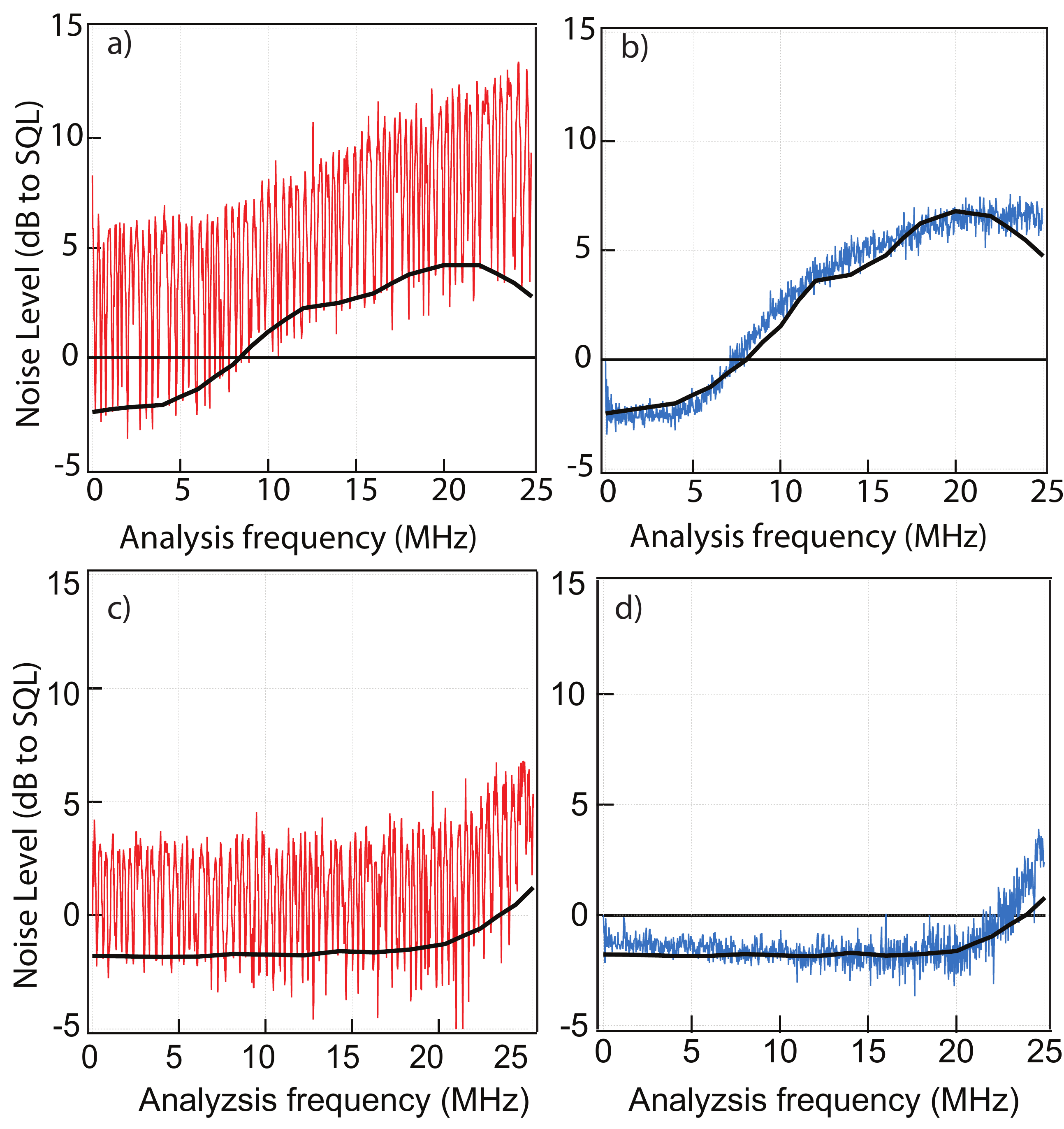}
\caption{\label{fig6} (Color online.)
Noise power as a function of the analysis frequency. 
Red traces (light gray) in a) and c) are identical to the red traces in Fig. 3 and correspond to the measured noise power when the phase of the LO is scanned. 
The black solid lines in a) and c) are $N_{min}(\omega_a,\delta)$.
The blue traces (dark gray) in b) and d) are identical to the blue traces in Fig. 3 and correspond to the measured noise power when the phase of the LO is locked for lowest noise at $1$~MHz.
The black solid lines in b) and d) are $N_{1}(\omega_a,\delta)$.
In a) and b) $\delta=-4$~MHz, and in c) and d) $\delta=-20$~MHz.
For the numerical simulations, we used  $\Delta \phi=0.2\pi$.}
\end{figure}

We have shown in the previous section (see Fig. \ref{fig4}) that we can compute $N(0,\delta-\omega_a,\phi_m)$ for $\phi_m$ that minimizes $N$.
Similarly, it is straightforward to compute this quantity for  $\phi_M$ that maximizes $N$.
We can then calculate the expected value of $N(0,\delta-\omega_a,\phi)$ for any value of $\phi$ :
 \begin{eqnarray}
	N(0,\delta-\omega_a,\phi)&=&N_-~ \text{Sin} (\phi) +N_+,
\end{eqnarray}	
	where 
	 \begin{eqnarray}
N_-=\frac{N(0,\delta-\omega_a,\phi_M)-N(0,\delta-\omega_a,\phi_m)}{2},\nonumber \\
 N_+=\frac{N(0,\delta-\omega_a,\phi_M)+N(0,\delta-\omega_a,\phi_m)}{2}. 
\end{eqnarray}
$\phi$ is chosen such that for $\phi=0$, we have $N(0,\delta-\omega_a,0)=N(0,\delta-\omega_a,\phi_M)$ and  for $\phi=\pi$, we have $N(0,\delta-\omega_a,\pi)=N(0,\delta-\omega_a,\phi_m)$.
Using these equations, we can compute $N_{min}(\omega_a,\delta)$ by minimizing the expression of Eq. (6) over $\phi$ and we can calculate $N_{1}(\omega_a,\delta)$.
The results of these simulations are plotted in Fig. \ref{fig6} for $\delta=-4$~MHz and $\delta=-20$~MHz, with $\Delta \phi=0.2\pi$.
The difference between $N_{min}$ and $N_{1}$, can then be explained by the presence of a phase shift between the two sidebands that contribute to the noise power spectrum.
The phase that minimizes the noise for $\omega_a=1$~MHz, is not necessarily the phase that minimize the noise at higher frequencies.
This effect can be interpreted as a rotation of the noise ellipse in phase space.
In the last section of this paper we propose a way to estimate the phase shift theoretically.

The model that we use is based on a microscopic model for a double-$\Lambda$ system, similar to the work of \cite{Glorieux:2010ja}.
Using the Heisenberg-Langevin formalism, we write the evolution equations for the atomic system and solve them analytically in the steady-state regime.
Using the slowly varying envelope approximation we can write the evolution of the probe field amplitude along the $z$ direction as:
	 \begin{eqnarray}
\frac{\text{d}}{\text{d}z} \begin{bmatrix}
  \alpha\\
\alpha^*
 \end{bmatrix}=\begin{bmatrix}
  A&B\\
B^*&A^*
 \end{bmatrix}\begin{bmatrix}
  \alpha\\
\alpha^*
 \end{bmatrix},
\end{eqnarray}
where $\alpha$ is the probe field amplitude and $A$ and $B$ are coefficients independent of $z$ that describe the steady state of the system.
$A$ and $B$  are calculated following a procedure similar to the derivation of Eq. (11) in  \cite{Glorieux:2010ja} and they depend on the pumps power, $\Delta$, $\delta$.
After introducing a term for the second pump field in Eq. (2) of \cite{Glorieux:2010ja}, we wrote the propagation equations similar to Eq. (6) and Eq. (7) of \cite{Glorieux:2010ja} for $\alpha$ and $\alpha^*$.
Finally, we can write $\alpha=\alpha_0\text{e}^{\text{i}\phi_\alpha}$, where $\alpha_0$ and $\phi_\alpha$ are real quantities.
The evolution equation for the phase $\phi_\alpha$ can then be written as: 
\begin{eqnarray}
\frac{\text{d}}{\text{d}z} \phi_\alpha=\text{Im}[A]+\text{Im}[B]\cos(2\phi_\alpha)-\text{Re}[B]\sin(2\phi_\alpha).\hspace{0.3cm}
\end{eqnarray}
We solve this equation and plot the value of $\phi_\alpha$ as function of the two-photon detuning in Fig. 7.
The numerical parameters used for the simulation are given in the figure caption and are directly calculated from the experimental values.
For this configuration we see that the phase shift is negligible for sidebands located bellow 0~MHz or above 20~MHz.
The value of the maximum observed phase shift (approximately 0.3$\pi$) is in agreement with the model we have developed in this paper. 
The physical origin of this phase shift lies in the susceptibility of the atomic medium, and that it depends on the two-photon detuning.
It is well known that in a three-level system in the $\Lambda$ configuration, the gain line around the Raman resonance, derived from the real part of the susceptibility, is coupled with a fast variation of the index of the medium derived from the imaginary part of the susceptibility.
The system we consider in this paper, presents the same qualitative behavior and we have observed experimentally the gain line appears around $\delta\approx15$~MHz.
Similarly, our theoretical model suggests a variation of the index (and in consequence a phase shift) that appears around $\delta\approx 10$ MHz, which is in qualitative agreement with the experimental results.

\begin{figure}[]
\centering\includegraphics[width=0.75\columnwidth]{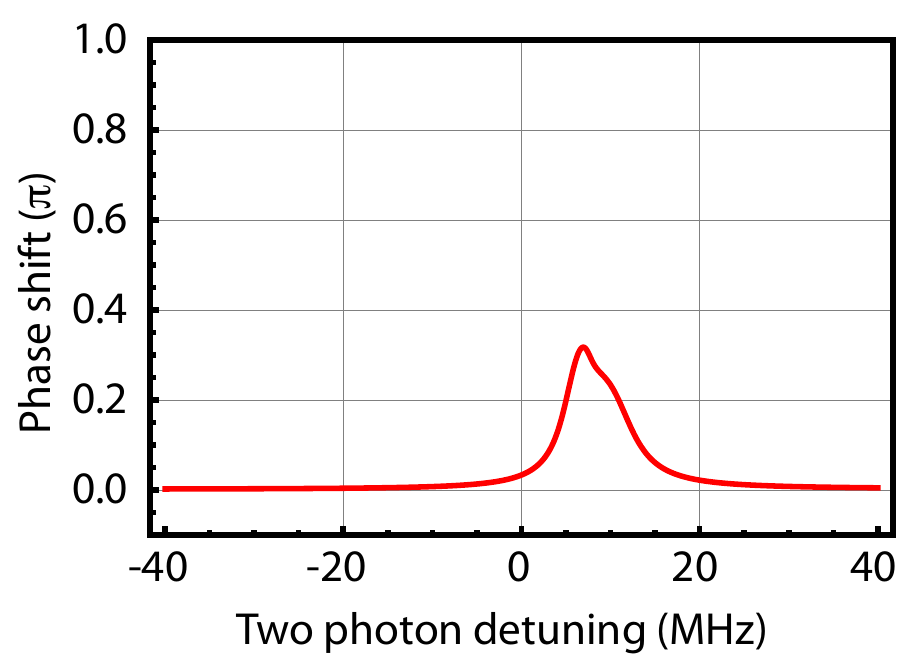}
\caption{\label{fig7} (Color online.) Phase shift as function of  $\delta$. 
Numerical simulation parameters are :   $\Delta=0.8$~GHz, pump power:~200~mW, optical depth:~1000.
}
\end{figure}

\section{Conclusion}

In this paper, we have presented the squeezing spectrum of the squeezed vacuum light generated via four-wave mixing in a double-$\Lambda$ system. 
We observe a rotation of the noise ellipse as function of different experimental parameters such as the detuning and pump power.
This rotation of the squeezing quadrature is of interest in applications such as gravitational wave detection because it could allow one to tailor the squeezed quadrature to match the dominant measurement noise at different frequencies, thus improving the noise performance across the spectrum.
Traditional techniques for generation of frequency-dependent squeezing rotation are based on the use of pre-filtering cavities.
Here we have shown that 4WM in a hot atomic vapor can provide this effect without the use of cavities.
An interesting feature of our system is the fact that it is spatially multimode and it has been shown that this property can be useful in sub-shot noise measurement techniques \cite{Treps:2002il}.
We use a theoretical model to describe the frequency dependent squeezing that we have observed and we suggest that this model can be used to investigate the parameter space (e.g. one-photon detuning, pump power, two-photon detuning) and to tailor the ellipse rotation in order to obtain the optimum squeezing angle rotation.

\acknowledgments{This work is supported by the AFSOR and the Physics Frontier Center of the Joint Quantum Institute. QG is supported by the Marie Curie IOF FP7 Program - (Multimem - 300632).}

\bibliography{biblio.bib}

\begin{thebibliography}{32}
\bibitem [{per()}]{perso}%
  \BibitemOpen
  \href@noop {} {\bibinfo  {journal} {The choice of the average rather than the
  sum here is arbitrary. Indeed, the noise power at $\omega_a=0$~Hz is a
  quantity not accessible experimentally and therefore can be defined with or
  without a factor of 2}\ }\BibitemShut {NoStop}%
\end{thebibliography}

\end{document}